\documentclass[twocolumn,times,tighten]{aastex631}

\usepackage{xspace}
\usepackage[T1]{fontenc}

\newcommand{\FeKa}{Fe K\ensuremath{\alpha}\xspace}

\newcommand{\kms}{\ensuremath{\mathrm{km\ s^{-1}}}\xspace}
\newcommand{\NH}{\ensuremath{N_{\mathrm{H}}}\xspace}

\newcommand{\slab}{\xspace{\tt slab}\xspace}

\newcommand{\xmm}{{XMM-Newton}\xspace}
\newcommand{\xrism}{{XRISM}\xspace}
\newcommand{\chandra}{{Chandra}\xspace}
\newcommand{\swift}{{Swift}\xspace}

\newcommand{\ergs}{{\ensuremath{\rm{erg\ s}^{-1}}}\xspace}
\newcommand{\cm}{{\ensuremath{\rm{cm}^{-2}}}\xspace}

\newcommand{\spex}{\xspace{\tt SPEX}\xspace}
\newcommand{\pion}{\xspace{\tt pion}\xspace}
\newcommand{\comt}{\xspace{\tt comt}\xspace}
\newcommand{\pow}{\xspace{\tt pow}\xspace}
\newcommand{\refl}{\xspace{\tt refl}\xspace}

\newcommand{\lya}{Ly\ensuremath{\alpha}\xspace}

\newcommand{\ngc}{{NGC~5548}\xspace}

\newcommand{\cf}{\ensuremath{C_f}\xspace}

\usepackage{xcolor}

\usepackage{xfrac}

\defcitealias{Mehd22c}{M22}

\newcommand{\MM}{{\citetalias{Mehd22c}}\xspace}
\received{2023 December 5}
\accepted{2024 January 5 by ApJ}
\shorttitle{First High-Resolution Spectroscopy of X-ray Absorption in the Obscured NGC 5548}
\shortauthors{Mehdipour et al.}
\graphicspath{{./}{figures/}}
\begin{document}

\title{\vspace{-0.2cm} \large First High-Resolution Spectroscopy of X-ray Absorption Lines in the Obscured State of NGC 5548}

\author[0000-0002-4992-4664]{Missagh Mehdipour}
\affiliation{Space Telescope Science Institute, 3700 San Martin Drive, Baltimore, MD 21218, USA; \href{mailto:mmehdipour@stsci.edu}{mmehdipour@stsci.edu}}

\author[0000-0002-2180-8266]{Gerard A. Kriss}
\affiliation{Space Telescope Science Institute, 3700 San Martin Drive, Baltimore, MD 21218, USA; \href{mailto:mmehdipour@stsci.edu}{mmehdipour@stsci.edu}}

\author[0000-0001-5540-2822]{Jelle S. Kaastra}
\affiliation{SRON Netherlands Institute for Space Research, Niels Bohrweg 4, 2333 CA Leiden, the Netherlands}
\affiliation{Leiden Observatory, Leiden University, PO Box 9513, 2300 RA Leiden, the Netherlands}

\author[0000-0001-8470-749X]{Elisa Costantini}
\affiliation{SRON Netherlands Institute for Space Research, Niels Bohrweg 4, 2333 CA Leiden, the Netherlands}
\affiliation{Anton Pannekoek Institute, University of Amsterdam, Postbus 94249, 1090 GE Amsterdam, The Netherlands}

\author[0000-0001-9911-7038]{Liyi Gu}
\affiliation{SRON Netherlands Institute for Space Research, Niels Bohrweg 4, 2333 CA Leiden, the Netherlands}
\affiliation{RIKEN High Energy Astrophysics Laboratory, 2-1 Hirosawa, Wako, Saitama 351-0198, Japan}
\affiliation{Leiden Observatory, Leiden University, PO Box 9513, 2300 RA Leiden, the Netherlands}

\author[0000-0001-8391-6900]{Hermine Landt}
\affiliation{Centre for Extragalactic Astronomy, Department of Physics, Durham University, South Road, Durham DH1 3LE, UK}

\author[0000-0001-7557-9713]{Junjie Mao}
\affiliation{Department of Astronomy, Tsinghua University, Haidian DS 100084, Beijing, People’s Republic of China}
\affiliation{SRON Netherlands Institute for Space Research, Niels Bohrweg 4, 2333 CA Leiden, the Netherlands}

\author[0000-0002-5359-9497]{Daniele Rogantini}
\affiliation{Department of Astronomy and Astrophysics, University of Chicago, Chicago, IL 60637, USA}
\affiliation{MIT Kavli Institute for Astrophysics and Space Research, Massachusetts Institute of Technology, Cambridge, MA 02139, USA}

\begin{abstract}

Multi-wavelength spectroscopy of NGC 5548 revealed remarkable changes due to presence of an obscuring wind from the accretion disk. This broadened our understanding of obscuration and outflows in AGN. Swift monitoring of NGC 5548 shows that over the last 10 years the obscuration has gradually declined. This provides a valuable opportunity for analyses that have not been feasible before because of too much obscuration. The lowered obscuration, together with the high energy spectral coverage of Chandra HETG, facilitate the first study of X-ray absorption lines in the obscured state. The comparison of the lines (\ion{Mg}{11}, \ion{Mg}{12}, \ion{Si}{13}, and \ion{Si}{14}) between the new and historical spectra reveals interesting changes, most notably the He-like absorption being significantly diminished in 2022. Our study finds that the changes are caused by an increase in both the ionization parameter and the column density of the warm-absorber outflow in the obscured state. This is contrary to the shielding scenario that is evident in the appearance of the UV lines, where the inner obscuring wind shields outflows that are located further out, thus lowering their ionization. The X-ray absorption lines in the HETG spectra appear to be unaffected by the obscuration. The results suggest that the shielding is complex since various components of the ionized outflow are impacted differently. We explore various possibilities for the variability behavior of the X-ray absorption lines and find that the orbital motion of a clumpy ionized outflow traversing our line of sight is the most likely explanation.

\end{abstract}
\section{Introduction} 
\label{sect_intro}

Understanding the co-evolution of active galactic nuclei (AGN) and their host galaxies has been one of the major endeavors of modern astronomy. AGN activity in the nucleus and star formation in the surrounding interstellar medium are thought to influence each other via a feedback mechanism \citep{Silk98,King15,Gasp17,Harr18}. AGN outflows/winds likely play a significant role as they serve as conduits of mass and energy into the host galaxy environment. To this end, finding out the uncertain properties of these outflows are important for assessing their contribution to AGN feedback.

The dynamics, kinematics, and ionization structure of ionized outflows, extending from the vicinity of the accretion disk to the outskirts of the host galaxy, are not well understood. This makes it challenging to ascertain how their momentum and energy propagate into the galaxy, and how they impact their environment. Different types/forms of ionized outflows, with distinct characteristics, have been observed at the micro (sub-pc) scale (disk and the broad-line region, BLR), the meso (pc) scale (the torus and the narrow-line region, NLR), and the macro (kpc) scale (the host galaxy environment), see e.g. \citet{Gasp20,Laha21}. The formation of these various ionized outflows, and their association to each other, are uncertain. The origin (disk or torus) and the launch and driving mechanism (thermal, radiative, or magnetic) of the ionized outflows remain open questions. Currently it is not well established what physical factors govern the launch and duty cycle of these winds.

The obscuring disk wind (obscurer) that was first discovered in \ngc \citep{Kaas14,Arav15} is significantly faster and more massive than the moderate ionized outflows (also called warm absorbers). While warm absorbers typically reside at pc-scale distances in the NLR, the obscurer is at distances of light days from the black hole, extending to and beyond the broad-line region (BLR). Such obscuring winds are transient, highly variable, and partially cover the central X-ray source \citep{DiGe15,Mehd16a,Capp16}. Apart from \ngc, they have been found in several other AGN, including Mrk 335 \citep{Long13}, NGC 985 \citep{Ebre16}, NGC 3783 \citep{Mehd17}, NGC 3227 \citep{Mehd21}, Mrk 817 \citep{Kara21}, PG 1114+445 \citep{Sera21}, and MR~2251-178 \citep{Mao22b}. Joint X-ray and UV spectroscopy has been instrumental in probing the nature of these obscuring winds. The appearance of broad and blueshifted UV absorption lines alongside X-ray obscuration in some objects suggests a link between these phenomena and enables the kinematics of the obscurer to be ascertained.

The HST/Cosmic Origins Spectrograph (COS) spectroscopy of \ngc suggests that the ionized outflows are shielded from the X-ray source by the obscurer, causing them to become less ionized \citep{Arav15,Kris19b}. This is seen by the detection of new low-ionization narrow absorption lines in the UV, such as \ion{C}{2} and \ion{Si}{2} \citep{Arav15,Mehd22c}. This scenario is also supported by the infrared spectroscopy of \ion{He}{1} \citep{Wild21} and the appearance of broad wings on some of the coronal emission lines \citep{Kyno22}. Interestingly, obscuration in \ngc is likely present in all directions, which would explain peculiar periods of de-correlation between the variabilities of the continuum and the BLR emission lines \citep{Dehg19a}. Therefore, such global obscuration/shielding impacts our interpretation of both the ionized outflows \citep{Dehg19b} and the reverberation mapping of the BLR \citep{Dehg19a}. Interestingly, shielding from the ionizing X-ray source is thought to be required to prevent over-ionization of the UV-absorbing gas and thus allowing radiative driving \citep{Prog04}. However, the need for this X-ray shielding in radiation-driven winds remains an open area of research in the literature (e.g. \citealt{Higg14}).
 
Recently in \citet{Mehd22c}, hereafter \MM, we reported on the striking long-term variability of the obscuring disk wind based on HST and \swift monitoring data (see Figs. 1 and 2 in \MM). The \swift/X-ray Telescope (XRT) hardness ratio is a useful tracer of the strength of X-ray obscuration. \swift showed that the X-ray spectral hardening in \ngc as a result of obscuration has declined over the years, reaching its lowest in 2022. At this point we found the broad \ion{C}{4} UV absorption significantly weakened, while the broad Ly$\alpha$ absorption was still significantly present. The associated narrow low-ionization UV absorption lines, produced previously by the warm absorber when shielded from the X-rays, were also remarkably diminished in 2022. We found a highly significant correlation between the variabilities of the X-ray spectral hardening and the equivalent width of the broad C IV absorption line, demonstrating that the X-ray obscuration is inherently linked to disk winds (\MM).

The lowered obscuration, and hence the higher signal-to-noise ratio (S/N) achieved with the brighter continuum, provides a new opportunity to study the X-ray absorption lines in the obscured state of \ngc. Here we present the high-resolution X-ray spectroscopy of the absorption lines with our new \chandra High Energy Transmission Grating (HETG) observations. This follows our study of the associated HST and \swift data in \MM. The ionized outflows of \ngc have been extensively studied in the historical unobscured epoch with high-resolution \xmm and \chandra spectra (e.g. \citealt{Kaas00,Stee05,Kaas14,Ebre16b}). The aim of this paper is thus not a full re-analysis of the ionized outflows, but rather to investigate any relative changes in them during the obscured epoch. This would help us better understand the relation between the obscurer and the ionized outflows, and ascertain the role and impact of shielding by the obscurer on the ionized outflows.

\section{Observations and data processing} 
\label{sect_obs}
\begin{figure}
\centering
\resizebox{1.0\hsize}{!}{\includegraphics[angle=270]{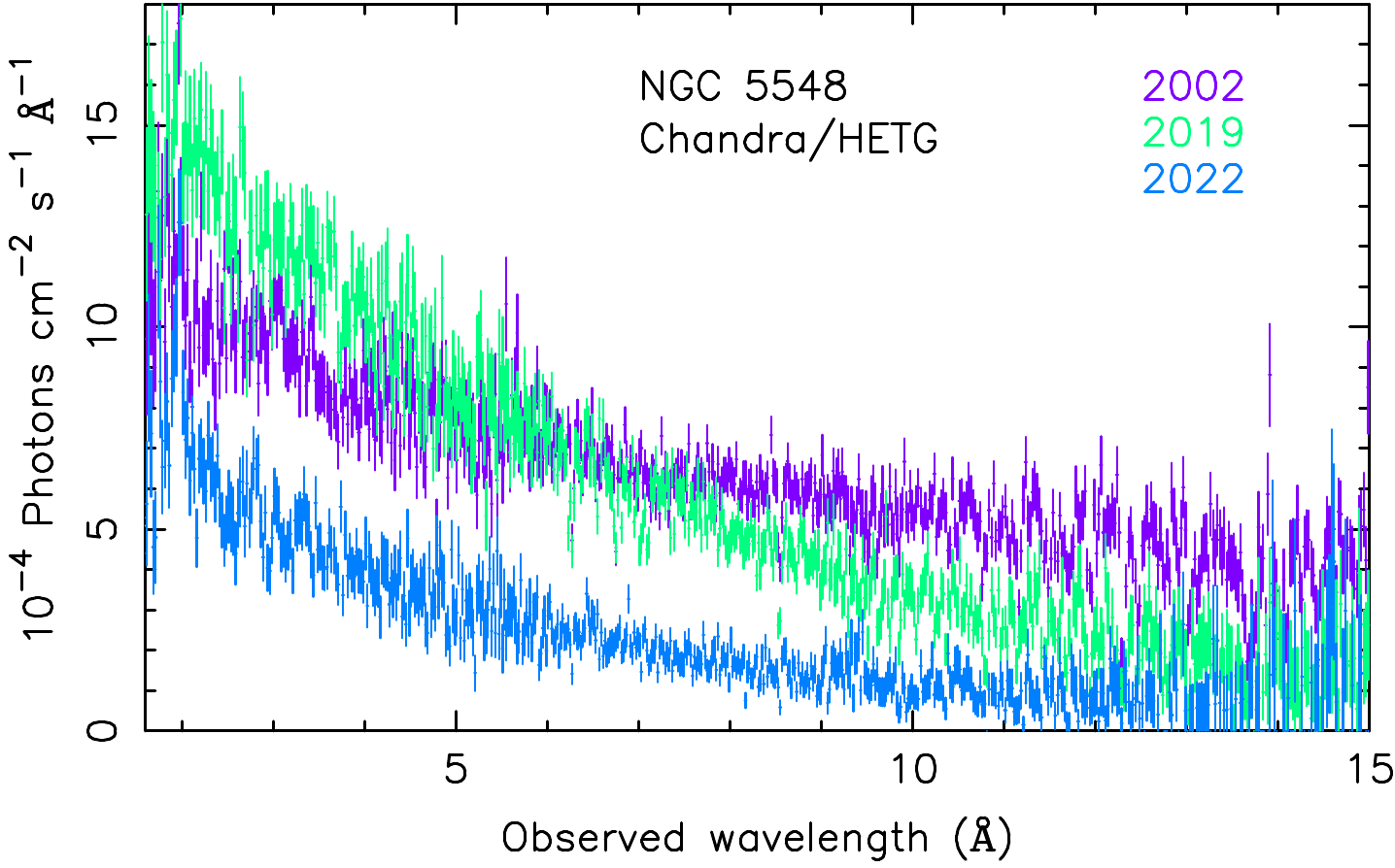}}
\caption{An overview of the new and historical HETG spectra of \ngc. The 2002 spectrum is from the unobscured epoch. The 2019 and 2022 spectra are both obscured, with the intrinsic continuum being brighter in 2019.
\label{fig_overview}}
\vspace{0.3cm}
\end{figure}

The log of the \chandra/HETG observations of \ngc that we are studying here are provided in Table \ref{table_log}. 
These \chandra data are contained in~\dataset[DOI: 10.25574/cdc.193]{https://doi.org/10.25574/cdc.193}.
An overview of the HETG spectra from the three epochs is shown in Figure \ref{fig_overview}.
Our most recent observation of \ngc was approved in a joint Chandra Cycle 23 proposal, providing a 150 ks exposure with HETG and 2 orbits with HST/COS.
To meet \chandra's observing requirement the observation was split, spanning December 2021 and January 2022. 
For brevity we refer to this as the ``2022'' observation.
Additionally, in 2019 a Guaranteed Time HETG observation with 175 ks exposure was obtained in the obscured epoch.
To help us better understand these 2019 and 2022 obscured spectra we make use of an unobscured observation from 2002 with 151 ks exposure.
This 2002 observation was accompanied with a HST/Space Telescope Imaging Spectrograph (STIS) observation.
In this paper we also make use of the 2013 HST/COS spectrum from the epoch of the strongest obscuration for comparison with other observations.
The new and archival HST spectra of \ngc have been described in our previous works (\citealt{Kaas14,Arav15,Kris19b}, \MM). Here we focus on the description and analysis of the HETG spectra.

%
\begin{deluxetable}{c c c c}[!t]
\tablecaption{Log of \chandra/HETG observations of \ngc from three epochs.
\label{table_log}}
\tablewidth{0pt}
\setlength{\tabcolsep}{10pt}
\tablehead{
Obs. ID & \colhead{Start Time} & \colhead{Exposure (ks)}
}
\startdata
3046    &       2002-01-16 06:12        & 151.4 \\
\hline
21846   &       2019-05-05 14:58        & 29.7 \\
22207   &       2019-06-18 13:16        & 50.3 \\
21694   &       2019-08-09 03:04        & 60.5 \\
22681   &       2019-08-10 22:39        & 27.2 \\
\hline
25802   &       2021-12-30 01:54        & 21.7 \\
26256   &       2021-12-30 15:16        & 21.7 \\
25803   &       2021-12-31 05:34        & 27.6 \\
25392   &       2022-01-30 18:31        & 24.7 \\
25800   &       2022-01-31 08:33        & 29.6 \\
25801   &       2022-01-31 22:18        & 19.8 \\
\enddata
\tablecomments{The individual HETG spectra from each epoch are combined into one spectrum for our modeling and are referred to as the ``2002'', ``2019'', and ``2022'' observations.}
\end{deluxetable}

In all the HETG observations, the ACIS camera was operated in the Timed Exposure (TE) read mode and the FAINT data mode. The data were reduced using the Chandra Interactive Analysis of Observations ({\tt CIAO}) v4.14 software. The {\tt chandra\_repro} script of {\tt CIAO} and its associated tools were used for the reduction of the data and production of the final grating products (PHA2 spectra, RMF and ARF response matrices). The grating spectra and their associated response files were combined using the CIAO {\tt combine\_grating\_spectra} script. The +/- first-order spectra of each grating were combined. In addition to producing a HEG and MEG spectrum for each observation, we also produced stacked HEG and MEG spectra containing all the data. We therefore produced three sets of spectra for our spectral modeling: 2002, 2019, and the 2022 spectra (Figure \ref{fig_overview}). The HETG spectra from each epoch display similar absorption features and are consistent with each other, thus allowing us to stack the spectra in order to enhance the S/N.  In our spectral modeling, HEG and MEG spectra are fitted simultaneously. The fitted spectral range is 2.5--26~\AA\ for MEG, and 1.55--14.5~\AA\ for HEG. We take into account the instrumental flux difference between HEG and MEG by re-scaling the normalization of HEG relative to MEG to in our spectral modeling. Over these energy bands, the HEG/MEG flux ratio is 0.96 in 2002, 0.94 in 2019, and 0.95 in 2022. 

\section{Spectral analysis and modeling} 
\label{sect_model}

We have carried out our spectral modeling using \spex v3.07.01 \citep{Kaa96,Kaas20}. In our modeling the Cosmological redshift is set to 0.017175 \citep{deVa91} using {\tt reds} in \spex. The X-ray absorption by the Milky Way is modeled with the {\tt hot} model with its temperature fixed to 0.001~eV and $N_{\mathrm{H}}={1.45\times 10^{20}\ \mathrm{cm}^{-2}}$ \citep{Wakk11}. The Galactic reddening is modeled using an {\tt ebv} component with ${E(B-V) = 0.02}$ \citep{Sch98} and $R_V$ fixed to 3.1

We modeled the HETG absorption lines with two different methods. In the first method we fitted each line using the \slab model. This allows the column density and velocity of an individual ion to be independently measured. In this approach prior knowledge of the ionizing spectral energy distribution (SED) and the ionization state of the gas is not needed, hence providing a more model agnostic diagnosis of the lines and their variability. In the second method, we use photoionization modeling using the {\pion} model \citep{Meh16b,Mill15}. This model takes into account the ionizing SED and computes the ionization balance of the gas, and thus all ionic column densities are linked in a physically consistent fashion. In our computations of the photoionization equilibrium and the X-ray spectrum, the elemental abundances are fixed to the proto-solar values of \citet{Lod09}.

%

%
\begin{figure*}
\centering
\resizebox{1.0\hsize}{!}{\includegraphics[angle=0]{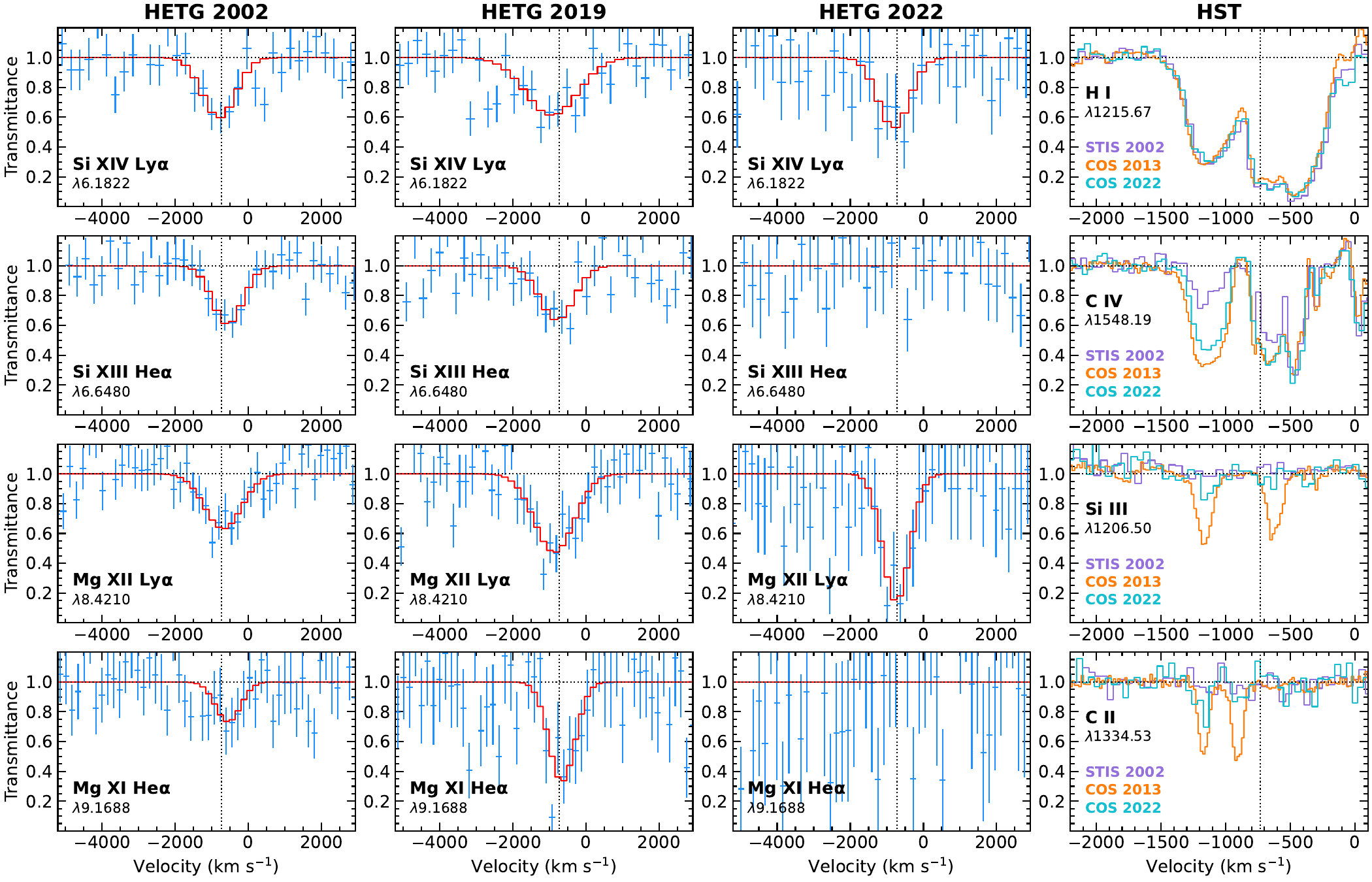}}
\caption{Profiles of the absorption lines in the high-resolution spectra of \ngc. The X-ray lines from the three epochs observed with HETG are compared in the first three columns. They are fitted with our \slab modeling shown in red. For comparison the profiles of the relevant UV absorption lines from HST observations are shown in the fourth column. The dotted vertical line in all panels is drawn at velocity of $-730$ \kms for visual reference.
\label{fig_lines}}
\vspace{0.9cm}
\end{figure*}

To model the broadband continuum we used the SED model that we have previously established for \ngc using our multi-wavelength campaign \citep{Meh15a}. In this model, the SED consists of three continuum components: \comt for modeling the optical/UV disk emission and the soft X-ray excess with warm Comptonization; \pow for modeling the X-ray power-law; \refl for modeling the X-ray reflection. The normalization and photon index of \pow are fitted to the HETG spectra. The low-energy and high-energy exponential cut-offs of \pow are fixed to 1 Ryd and 400~keV, respectively \citep{Meh15a}. We scale the normalization of \comt model to match the HST UV continuum level. The other parameters of \comt are kept frozen to those obtained from the 2013 campaign \citep{Meh15a} since they cannot be constrained with HETG, mainly because of the loss of effective area in the soft band. Also, due to lack of any UV data for the 2019 observation we set the normalization of its \comt to that of the 2022 observation. The parameters of the illuminating power-law for the \refl component are coupled to those of the 2002 intrinsic power-law, and the scale parameter of the \refl model is freed to fit the \FeKa line.

Despite the gradual long-term decline of the obscurer, it still significantly absorbs the spectrum (\MM). Therefore, we need to take into account its presence in our modeling of the HETG spectra. To fit the obscurer we adopt the model of \citet{Kaas14} that was derived from the large \xmm campaign. In this model the obscurer has two components, representing denser (colder) clumps embedded in a more diffuse (warmer) medium. Previous studies have found the covering fraction (\cf) of the warm phase is the main variable parameter \citep{Mehd16a,Capp16}. As discussed in \MM, the decline in obscuration is attributed to lowering of the covering fraction of the obscurer. Therefore, we fit \cf in our modeling of the HETG spectra. Other parameters of the obscurer are kept fixed as they are not needed to be re-fitted and also it would be challenging to constrain them with \chandra alone.

\begin{figure}[t]
\centering
\resizebox{1.0\hsize}{!}{\includegraphics[angle=0]{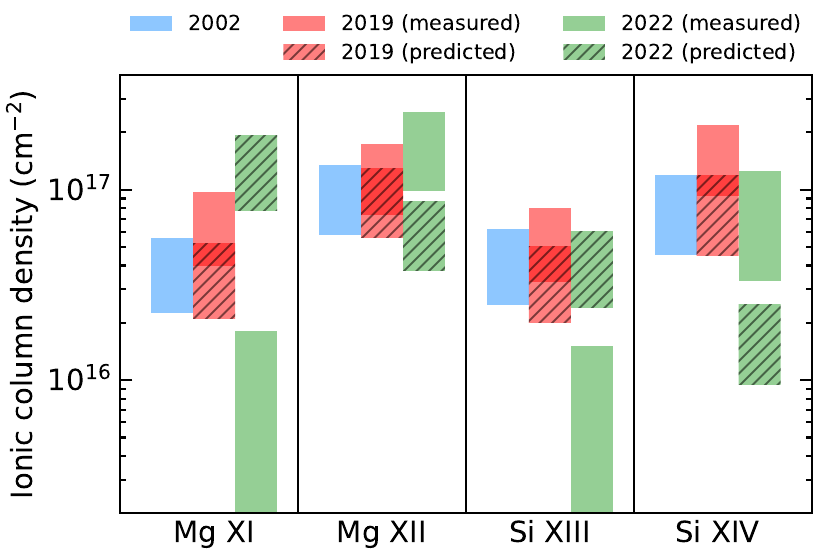}}
\caption{Column densities of individual ions measured via \slab modeling of absorption lines in the HETG spectra of the three epochs. The height of each band represents the measurement uncertainty. The ``predicted'' values for the 2019 and 2022 epochs correspond to the de-ionization scenario, where the 2002 outflow is shielded by the obscurer and thus is illuminated by the obscured SED. In the 2022 observation the measured and predicted ionic column densities differ significantly.
\label{fig_col}}
\vspace{0.3cm}
\end{figure}

\begin{figure*}
\centering
\resizebox{1.0\hsize}{!}{\includegraphics[angle=270]{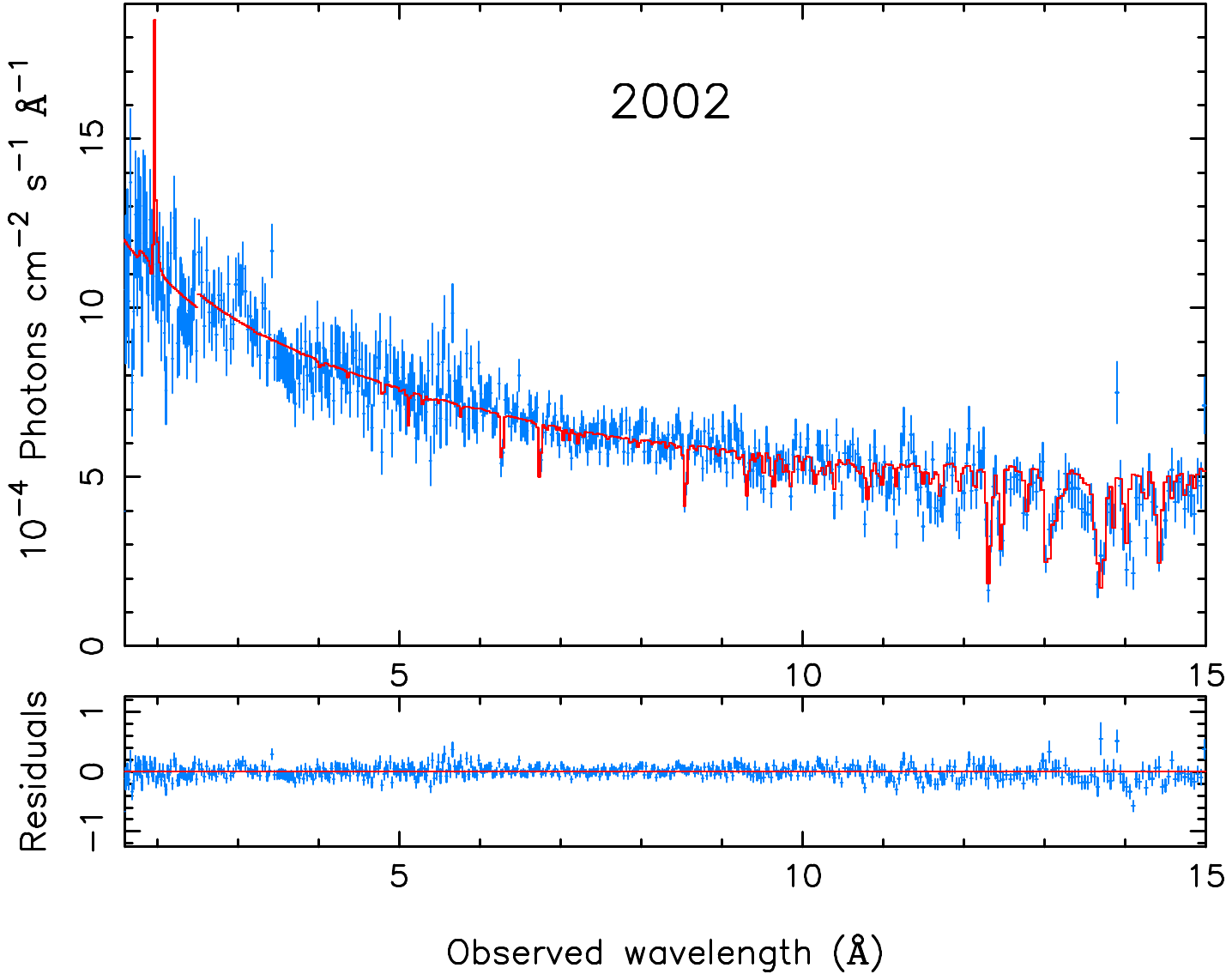}\includegraphics[angle=270]{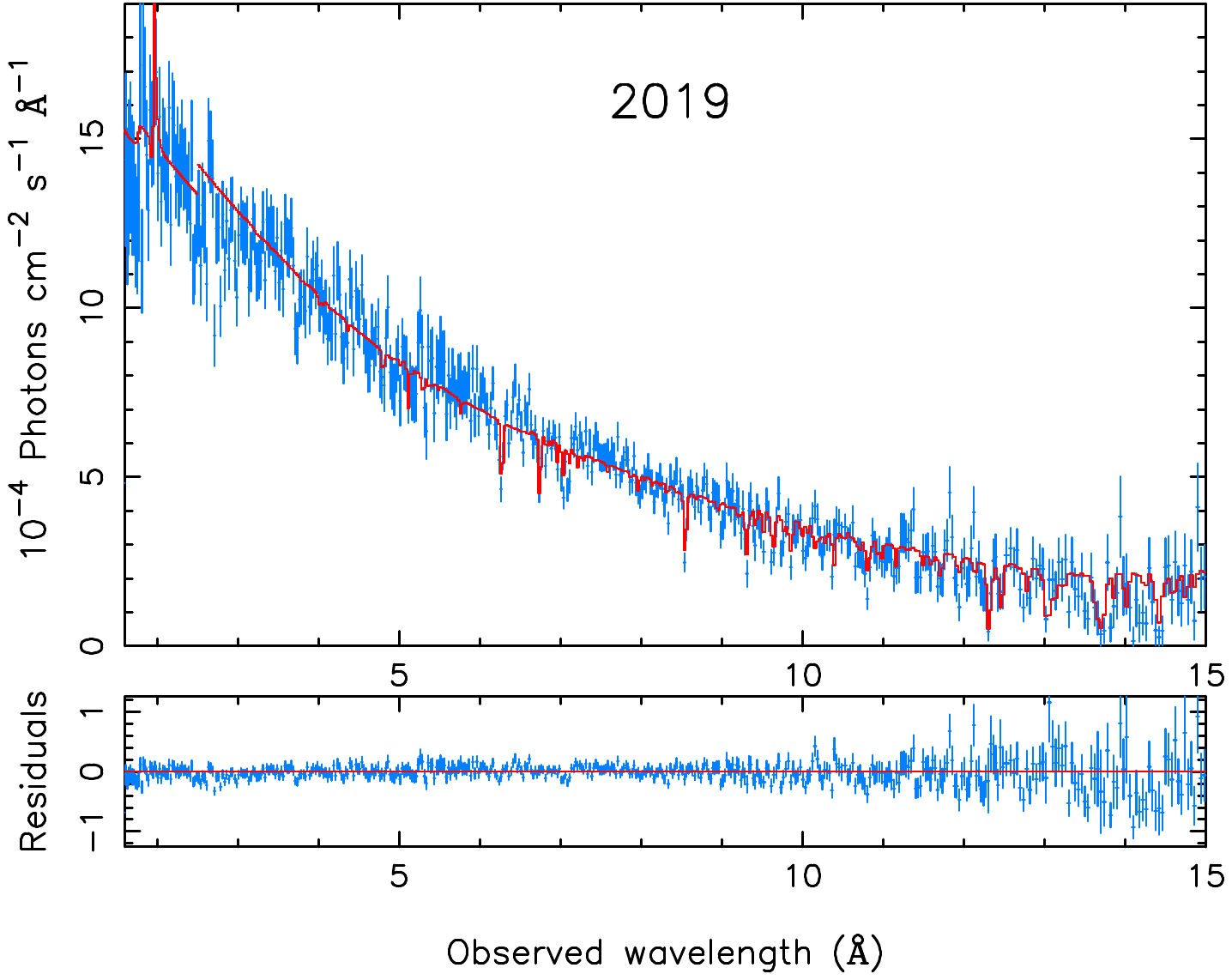}\includegraphics[angle=270]{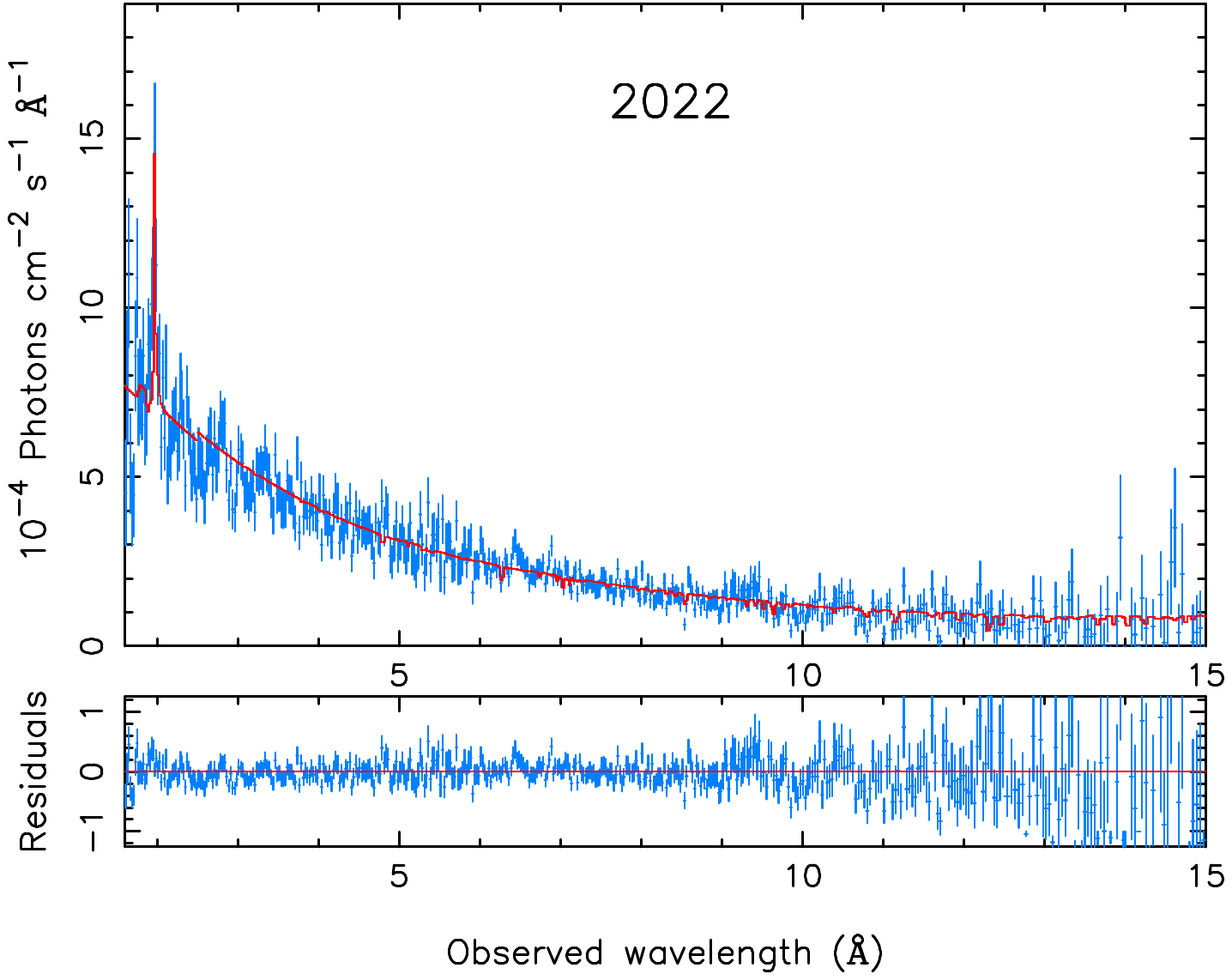}}
\caption{Best-fit \pion model to the 2002 (left panel), 2019 (middle panel), and 2022 (right panel) HETG spectra of \ngc. Residuals of the fits, defined as (data $-$ model) / model, are displayed in the bottom panels. The parameters of the \pion photoionization modeling are given in Table \ref{table_para} and the associated SED models are displayed in Figure \ref{fig_sed}.
\label{fig_fit}}
\vspace{0.5cm}
\end{figure*}

Figure \ref{fig_lines} shows the absorption lines that are significantly detected with HETG in the 2019 and 2022 obscured epochs. They are compared with the unobscured HETG spectrum from 2002, as well as the absorption profiles of the key UV lines from HST observations. Our measured ionic column densities using the \slab modeling are displayed in Figure \ref{fig_col}. In this figure we compare what is measured from observation with the ``predicted'' one, which would correspond to the 2002 gas only responding to the ionizing SED in 2019 and 2022 (i.e. the de-ionization scenario due to shielding). The best-fit model to the HETG spectra using \pion modeling are shown in Figure \ref{fig_fit}, and the corresponding parameters are given in Table \ref{table_para}. The associated SED models that we have derived for each epoch using our continuum modeling (\comt, \pow, and \refl components) are shown in Figure \ref{fig_sed}. The unobscured SEDs show the intrinsic continuum while the obscured SEDs include absorption by the obscurer. The corresponding luminosities of the SEDs are reported in Table \ref{table_lum}. We discuss these results in the following section.

%

\section{Discussion and conclusions} 
\label{sect_discuss}

In this paper we investigated how the ionized outflow in \ngc has evolved and assessed the impact of the obscurer on the ionized outflow. 
The decline in obscuration, and the high-energy spectral coverage of HETG, make it possible to carry out high-resolution X-ray spectroscopy of the absorption lines to measure any changes in the parameters of the ionized outflow.
The previous X-ray studies of \ngc in the historical unobscured epoch have shown that its ionized outflow consists of multiple ionization and velocity components \citep{Kaas14,Ebre16b}. 
However, \chandra's loss of effective area in the soft X-ray band due to the ACIS contamination means that only the high-ionization component of the outflow can be constrained in the 2019 and 2022 obscured spectra.
The detection of other components at longer wavelengths is nonetheless challenging due to still significant absorption of the continuum by the obscurer.
The absorption lines that are most significantly detected with HETG belong to \ion{Mg}{11}, \ion{Mg}{12}, \ion{Si}{13}, and \ion{Si}{14} as shown in Figure \ref{fig_lines}.
We find interesting changes in the strength of these lines that we interpret here based on the results of our modeling. 

\subsection{Impact of the obscurer on the UV and X-ray components of the ionized outflows}

Evidence for shielding (de-ionization) of the ionized outflows by the obscurer is seen in the HST UV spectra of \ngc. 
This was most recently discussed in \MM, where the strength of the low-ionization absorption lines follows the strength of the X-ray obscuration.
This effect can be seen in the appearance of the \ion{C}{2} and \ion{Si}{3} lines in Figure \ref{fig_lines}, where they are strongest in 2013 (i.e. the epoch of strongest X-ray obscuration) and have become significantly weaker in 2022 (due to the lowered obscuration). In 2002 these lines are not detected as \ngc was unobscured.
Interestingly, in our HETG study we find that the ionized outflow in X-rays does not follow the de-ionization scenario that is apparent in the UV.
The results of both types of our modeling, \slab (independent fitting of individual lines) and \pion (photoionization modeling), are in agreement and support this finding.

Figure \ref{fig_col} shows that the ionic column density of the He-like species (\ion{Mg}{11} and \ion{Si}{13}) became significantly lower in 2022.
As shown in Table \ref{table_para} this is caused by an increase in the ionization parameter $\xi$ and the column density \NH of the ionized outflow compared to the 2002 unobscured epoch.
This behavior is contrary to the de-ionization scenario where the shielding by the obscurer would lower the ionization of the gas.
Interestingly, in the case of \ion{C}{4} absorption at around $-800$~\kms in Figure \ref{fig_lines} (i.e. close to the outflow velocity of the ionized outflow that is seen in X-rays) the absorption in 2022 is weaker than in 2013.
This is most likely because the \ion{C}{4} absorbing gas became more ionized, like seen in the X-rays by HETG.
It is worth noting that the \ion{H}{1} \lya line appears unchanged in Figure \ref{fig_lines} because this partially-covering feature is highly saturated, thus changes in the \ion{H}{1} column density cannot produce any discernible variability in the absorption profile.

%
\begin{deluxetable}{c | c c c}
\tablecaption{Best-fit parameters of the ionized outflow (warm absorber), the obscurer, and the continuum components from \pion modeling of the three HETG spectra.
\label{table_para}}
\tablewidth{0pt}
\setlength{\tabcolsep}{5pt}
\tablehead{
Parameter & \colhead{2002} & \colhead{2019} & \colhead{2022}
}
\startdata
Ionized outflow: \\
$\log~\xi$ (erg~cm~s$^{-1}$)    & $2.46 \pm 0.02$  & $2.54 \pm 0.03$  &  $2.96 \pm 0.05$  \\
\NH ($10^{21}$ \cm)             & $4.0 \pm 0.3$    & $5.3  \pm 0.3$   &  $8.0 \pm 1.0$   \\
$v_{\rm out}$ (\kms)            & $730 \pm 20$    & $730$ (c)       &  $730$ (c)        \\
$\sigma_v$ (\kms)               & $390 \pm 30$     & $390$ (c)        &  $390$ (c)        \\
\hline
Obscurer: \\
\cf                             & 0 (f)            & $0.75  \pm 0.03$   &  $0.65 \pm 0.05$    \\
\hline
Continuum: \\
\pow Norm.                      & $4.8 \pm 0.1$    &  $10.3 \pm 0.1$    & $3.1 \pm 0.1$                  \\
\pow $\Gamma$                   & $1.64 \pm 0.02$  &  $1.84 \pm 0.03$   & $1.61 \pm 0.03$                  \\
\comt Norm.                     & $3.6$ (f)        &  $6.0$ (f)         & $6.0$ (f)                  \\
\refl scale                     & $0.37 \pm 0.04$  &  $0.37$ (c)        & $0.37$ (c)                   \\
\hline
C-stat\,/ d.o.f.			& 5613\,/\,5122 & 5173\,/\,4589		&		4581\,/\,4158	\\
\enddata
\tablecomments{The normalization of the power-law component (\pow) is in units of $10^{51}$ photons~s$^{-1}$~keV$^{-1}$ at 1 keV, and the normalization of the Comptonization component ({\tt comt}) is in $10^{55}$ photons~s$^{-1}$~keV$^{-1}$. The ``(f)'' denotes that the value of the parameter is kept fixed and ``(c)'' means the parameter is coupled to another one in our modeling.}
\end{deluxetable}

%

\begin{figure*}
\centering
\resizebox{1.0\hsize}{!}{\includegraphics[angle=0]{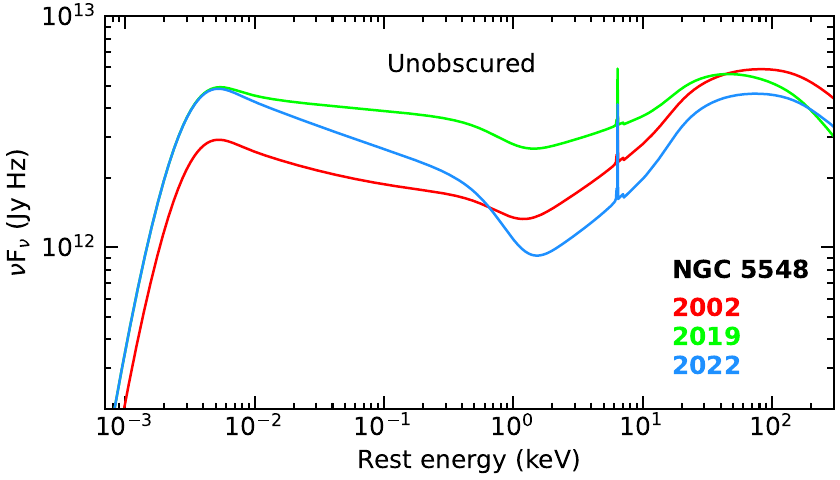}\includegraphics[angle=0]{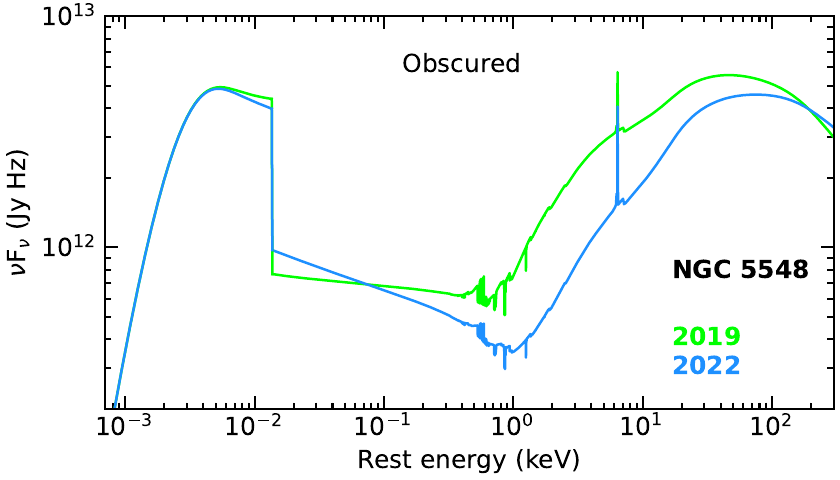}}
\caption{Derived SED models for the three epochs of \ngc. The best-fit parameters of the continuum components are provided in Table \ref{table_para}. The unobscured SEDs, shown on the left panel, would illuminate the obscurer. The obscured SEDs, shown on the right panel, would illuminate any outflow that is located further out and shielded by the obscurer. The luminosities of these unobscured and obscured SEDs are given in Table \ref{table_lum}.
\vspace{0.5cm}
\label{fig_sed}}
\end{figure*}

The results of our SED modeling suggest that change in the intrinsic broadband continuum is not responsible for the observed variability of the HETG absorption lines. We have derived the SED for each epoch (Figure \ref{fig_sed} and Table \ref{table_para}) and carried out photoionization modeling. The observed change in the shape and luminosity of the SED (Figure \ref{fig_sed} and Table \ref{table_lum}) is not sufficient to induce the significant change in the ionization parameter $\xi$ of the ionized outflow (Table \ref{table_para}). The parameter $\xi$ is higher by a factor of about three in 2022 than in 2002, while the luminosity of the ionizing continuum is only 10\% higher. Therefore, regardless of whether the unobscured or obscured SED illuminates the ionized outflow, the observed change in the X-ray absorption lines cannot be explained by the SED variability. We note that this HETG result is also supported by the \swift X-ray and UV monitoring of \ngc (\MM), which shows no significant flaring occurred near or during our observations to induce such a jump in the ionization of the gas. Interestingly, the X-ray line variability seen by HETG is in contrast to the general variability of the UV absorption lines, which predominantly follow the ionizing SED \citep{Arav15}.

The best-fit parameters of Table \ref{table_para} also indicate that the total column density \NH of the warm absorber increases with the ionization parameter $\xi$ over the three epochs that we have investigated.
Such a trend was previously found by \citet{Stee03} for the historical warm absorber of \ngc where \NH versus $\xi$ was seen to follow a power-law distribution.
This relation is thought to be a manifestation of the optical depth of the absorber being constant \citep{Dehg21}.

The ionization energies for the production of the two He-like ions that have disappeared in 2022 (\ion{Mg}{11} and \ion{Si}{13}) are in the range of 0.4--0.5 keV. This corresponds to the energy band where the ionizing SED that is transmitted by the SED is significantly diminished as shown by the computations of \citet{Dehg21}. On the other hand, the ionization energies for the production of the two H-like ions (\ion{Mg}{12} and \ion{Si}{14}) are in the higher range of 1.8--2.4 keV (Table \ref{table_lum}), which is less affected by the obscuration.  
Since the obscurer absorbs a significant amount of the incident SED, it would also re-emit to conserve energy \citep{Dehg21}.  
This might imply that in 2022 the warm absorber was not illuminated by the reprocessed emission from the obscurer and was only seeing the transmitted SED, which is diminished over the 0.4--0.5 keV band.
However, in the 2019 observation the \ion{Mg}{11} and \ion{Si}{13} lines are detected, which would imply that the warm absorber was illuminated by the reprocessed emission.
The origin of this behavior between 2019 and 2022 is uncertain, but may be due to changes in the parameters of the obscurer, in particular its ionization parameter, which would alter the spectral characteristics of its reprocessed emission.
However, measuring the ionization parameter of the obscurer and its changes has been extremely challenging in all previous studies and cannot be obtained from fitting the HETG spectra.
As discussed in \citet{Kris19b}, due to the complex and multi-component nature of the obscurer, finding a unique photoionization solution has not been feasible.

%

\subsection{Interpretation of the variability of the HETG X-ray absorption lines}

The study of historical unobscured spectra by \citet{Ebre16b} derived constraints for the location of various components of the ionized outflows in \ngc. Their component E is the one that matches most closely our model in terms of $\xi$ and outflow velocity. They found the distance of this component from the black hole is ${0.6 < R < 2.2}$~pc. We suggest the obscuring wind itself is unlikely to have reached such distances to make a direct contribution to the observed HETG absorption lines. The earliest available evidence for the obscurer comes from the \swift observation of February 2012 \citep{Mehd16a} in the X-ray band and the HST observation of June 2011 in the UV band \citep{Kaas14}. This is about 11 years before our 2022 HETG observation. The precise time for a potentially earlier appearance of the obscurer is not known due to lack of regular monitoring prior to 2012, however, the August 2007 \swift observation is unobscured \citep{Mehd16a}. Therefore, the maximum possible travel time for the obscurer before our 2022 HETG observation is about 14.5 years. The outflow velocity of the obscurer according to the broad UV absorption lines is dominant at around 2000 \kms, reaching up to 5000 \kms \citep{Kaas14,Kris19b}. With this velocity range, the obscurer would have traveled a distance of 27--67 light days (0.02--0.06 pc) over 11 years, or potentially a maximum distance of 35--88 light days (0.03--0.07 pc) over 14.5 years. These distances correspond to the outer BLR rather than the NLR and the estimated location of ${0.6 < R < 2.2}$~pc by \citet{Ebre16b}.

The HETG absorption lines are significantly narrower (few 100 \kms) than the broad UV lines (few 1000 \kms) that are associated with the obscurer in the vicinity of the BLR. Also, the outflow velocity of the HETG absorption lines shows no significant change between the three epochs, again suggesting that the obscurer is not directly contributing to these lines. The absorption lines in the 2002 and 2019 epochs have similar strengths, whereas they show key differences between 2019 and 2022 (Figure \ref{fig_lines}). This points to variability on shorter timescales than the decadal timescale of the evolution of the obscurer.

The most plausible explanation for the observed line variability may be due to the orbital motion of the warm-absorber outflow as it traverses our line of sight. At distance of ${0.6 < R < 2.2}$~pc \citep{Ebre16b}, the Keplerian velocity would range from 370 to 710 \kms for \ngc's black hole mass of $7 \times 10^7$~$M_{\odot}$ \citep{Horn21}. This orbital velocity range is consistent with our measured velocity dispersion of 390 \kms, as well as with the outflow velocity of 730 \kms (Table \ref{table_para} and Figure \ref{fig_lines}). Adopting a fiducial X-ray source size (diameter) of 20 gravitational radii ${GM/c^{2}}$, the crossing time with the Keplerian velocity would be 15--30 days. Therefore, on the longer timescales that we are probing with our three observations (2002, 2019, and 2022), the absorbing gas can feasibly be replaced by new material. In the likely case of this gas being inhomogeneous and clumpy, as it moves, variations in the ionization parameter and the column density are expected, like seen in our HETG observations. 

%
\begin{deluxetable}{c | c c c}
\tablecaption{Luminosities of the unobscured and obscured SEDs of \ngc (Figure \ref{fig_sed}) for the three epochs.
\label{table_lum}}
\tablewidth{0pt}
\setlength{\tabcolsep}{11pt}
\tablehead{
 & \colhead{2002} & \colhead{2019} & \colhead{2022}
}
\startdata
Unobscured SED: \\
$L_{\rm 1-1000}$    &  0.9 &  1.6 &		1.0	\\
$L_{\rm bol}$		 &  2.5  &  3.4 &	  2.6		\\
\hline
Obscured SED: \\
$L_{\rm 1-1000}$    &  -  &  0.6  &		0.4	\\
$L_{\rm bol}$		 &  -  & 	2.3   &   2.0		\\
\hline
$L_\nu$ ratio at ionization potential:\\
\ion{H}{1} (13.60 eV)     &   -   &   0.17  &    0.25  \\
\ion{C}{2} (24.38 eV)     &   -   &   0.17  &    0.25  \\
\ion{C}{4} (64.49 eV)     &   -   &   0.17  &    0.25  \\
\ion{Mg}{11} (1761.80 eV) &   -   &   0.52  &    0.54  \\
\ion{Mg}{12} (1962.66 eV) &   -   &   0.54  &    0.57  \\
\ion{Si}{3} (33.49 eV)    &   -   &   0.17  &    0.25  \\
\ion{Si}{13} (2437.66 eV) &   -   &   0.64  &    0.66  \\
\ion{Si}{14} (2673.18 eV) &   -   &   0.67  &    0.68  \\
\enddata
\tablecomments{$L_{1-1000}$ corresponds to the 1--1000 Ryd ionizing luminosity and $L_{\rm bol}$ to the total bolometric luminosity of the SED. The luminosities are in units of $10^{44}$ \ergs. The $L_\nu$ ratio corresponds to the obscured versus unobscured specific luminosity ratio at the energy of the ionization potential of an ion.}
\vspace{-0.3cm}
\end{deluxetable}

The results of this study highlight the complex relation between the obscurer and the ionized outflows in \ngc. While some of the more distant outflows in the UV appear to be shielded, the more ionized outflow that is seen in X-rays with HETG is not significantly affected by the obscurer.
In addition, the shielding has a greater impact on the population of the low-ionization UV ions that is more dramatic than the high energy species seen with the HETG due to the higher optical depth of the obscurer at their relevant ionization edges. This can be seen by the obscured versus unobscured specific luminosity ratio at the ionization potential of each ion (Table \ref{table_lum}).
It is not certain to what extent the origin and geometry of the obscurer and the warm-absorber outflow play a role. For instance, if the obscurer is launched from the disk and mainly moves in polar directions as a result of being magnetically driven \citep{Fuku17}, it might not effectively shield regions in the equatorial directions, where the warm-absorber outflow (Component E of \citealt{Ebre16b}) may originate with material evaporated from the torus. Thus, in such a case the obscurer and the ionized outflow would operate and vary independently from each other. On the other hand, if the obscurer and the warm-absorber outflow share a common origin and launching mechanism, one would expect some degree of observable correlation between their variability behaviors.
Furthermore, the variable and patchy nature of the obscurer may play a role by letting through various ionizing SEDs that would illuminate the warm absorber, causing short term changes in the ionization of the warm absorber as seen in the 2019 and 2022 epochs.

The potential role of X-ray shielding in radiation-driven winds is open for further research. The hydrodynamical simulation study of \citet{Higg14} finds that the shielding may not be effective in keeping the outer UV absorber from being over-ionized because of reprocessing and scattering of the ionizing X-rays. Also, the study of broad absorption line quasars by \citet{Luo14} suggests that they are intrinsically X-ray faint and do not need shielding to drive winds. With HETG we are only able to probe the highest ionization component of the warm-absorber outflow and our findings suggest that it is not significantly shielded by the obscurer. The recently launched \xrism/Resolve microcalorimeter \citep{xrism20}, alongside \xmm/RGS, present a unique opportunity to probe all the X-ray components of the outflow in \ngc as obscuration further evolves over the coming years.

\begin{acknowledgments}
Support for this work was provided by the National Aeronautics and Space Administration through Chandra Award Number 23103X issued by the Chandra X-ray Center, which is operated by the Smithsonian Astrophysical Observatory for and on behalf of the National Aeronautics and Space Administration under contract NAS8-03060. This work was also supported by NASA through a grant for HST program number 16842 from the Space Telescope Science Institute, which is operated by the Association of Universities for Research in Astronomy, Incorporated, under NASA contract NAS5-26555. SRON is supported financially by NWO, the Netherlands Organization for Scientific Research. We thank the anonymous referee for the constructive comments.
\end{acknowledgments}
\facilities{Chandra (HETG), HST}
\bibliographystyle{aasjournal}
\bibliography{references}{}
\end{document}